\documentclass[paper,notoc,nohyper]{JHEP}
\usepackage{amsmath}
\usepackage{epsfig}

\def\ltap{\raisebox{-.4ex}{\rlap{$\,\sim\,$}} \raisebox{.4ex}{$\,<\,$}}
\def\gtap{\raisebox{-.4ex}{\rlap{$\,\sim\,$}} \raisebox{.4ex}{$\,>\,$}}

\newcommand\as{\alpha_{\mathrm{S}}}
\newcommand\f[2]{\frac{#1}{#2}}
\def\dO{{\cal D}_{0}}
\def\dl{{\cal D}_{1}}
\def\dll{{\cal D}_{2}}
\def\dlll{{\cal D}_{3}}

\def\beq{\begin{equation}}
\def\eeq{\end{equation}}
\def\beeq{\begin{eqnarray}}
\def\eeeq{\end{eqnarray}}

\def\to{\rightarrow}
\def\nn{\nonumber}

\def\ms{${\overline {\rm MS}}$}
\def\msbar{{\overline {\rm MS}}}
\def\asp{\f{\alpha_s}{\pi}}

\renewcommand{\thefootnote}{\fnsymbol{footnote}}

\preprint{CERN--TH/2001-044}

\title{\boldmath Higgs production in hadron collisions: soft and virtual QCD corrections at NNLO
\footnote{This work was supported in part 
by the EU Fourth Framework Programme ``Training and Mobility of Researchers'', 
Network ``Quantum Chromodynamics and the Deep Structure of
Elementary Particles'', contract FMRX--CT98--0194 (DG 12 -- MIHT).}
\\[1.ex]
}

\author{Stefano Catani${}^{(a)}$~\footnote{On leave of absence 
from INFN, Sezione di Firenze, Florence, Italy.}, Daniel de Florian${}^{(b)}$
\footnote{Partially supported by Fundaci\'on Antorchas}
and
Massimiliano Grazzini${}^{(c,d)}$\\
${}^{(a)}$Theory Division, CERN, CH-1211 Geneva 23, Switzerland \\[1mm]
${}^{(b)}$Theoretical Physics, ETH-H\"onggerberg, 
CH-8093 Zurich, Switzerland \\[1mm]
${}^{(c)}$Dipartimento di Fisica, Universit\`a di Firenze, I-50125 Florence, Italy\\[1mm]
${}^{(d)}$INFN, Sezione di Firenze, I-50125 Florence, Italy
}

\abstract{
We consider QCD corrections to Higgs boson production through gluon--gluon
fusion in hadron collisions. 
Using the recently evaluated [14]
two-loop amplitude for this process and the corresponding factorization
formulae [15, 16, 17, 18] 
describing soft-gluon bremsstrahlung at ${\cal O}(\as^2)$, we
compute the soft and virtual contributions to the 
next-to-next-to-leading order cross section.
We also discuss soft-gluon resummation at next-to-next-to-leading
logarithmic accuracy. Numerical results for Higgs boson production
at the LHC are presented.
}

\begin{document}

\setcounter{footnote}{1}
\renewcommand{\thefootnote}{\fnsymbol{footnote}}


\section{Introduction}
\label{sec:intro}

The Higgs boson \cite{Gunion:1989we}
is a fundamental ingredient of the Standard Model (SM),
but it has not yet been observed.

Direct searches at LEP imply a lower limit of $M_H > 112.3$~GeV (at $95\%$ CL) 
\cite{Junk:2001ad} on the mass $M_H$ of the SM Higgs boson.
Global SM fits to electroweak precision measurements favour a light Higgs
($M_H \ltap 200$~GeV) \cite{Wynhoff:2001cg}.
The combination of the preliminary Higgs boson search results of the four LEP
experiments \cite{lepc,leppapers} shows an excess of candidates, which may
indicate the production of a SM Higgs boson with a mass near 115~GeV.
The final analysis of the LEP data is expected soon, but it is unlikely that it
can substantially change these results.

After the end of the LEP physics programme, the search for the Higgs boson 
will be carried out at hadron colliders. Depending on the luminosity
delivered to the CDF and D0 detectors during the forthcoming Run II, the 
Tevatron experiments can yield evidence for a Higgs boson with 
$M_H < 180$~GeV and may be able to discover (at the $5\sigma$ level)
a Higgs boson with $M_H \ltap 130$~GeV \cite{Carena:2000yx}.
At the LHC, the SM Higgs boson can be discovered over the full mass range up
to $M_H \sim 1$~TeV after a few years of running \cite{atlascms}.

The dominant mechanism for SM Higgs boson production at hadron colliders is
gluon--gluon fusion through a heavy-quark (top-quark) loop \cite{Georgi:1978gs}.
At the Tevatron, this production mechanism leads to about $65\%$ of the total
cross section for producing a Higgs boson in the mass range $M_H=100$-200~GeV
\cite{Carena:2000yx}. At the LHC \cite{Spira:1998dg}, 
$gg$ fusion exceeds all the other production
channels by a factor decreasing from 8 to 5 when $M_H$ increases from 100 to
200~GeV. When $M_H$ approaches 1~TeV, $gg$ fusion still provides about 
$50\%$ of the total production cross section.

QCD radiative corrections at next-to-leading order (NLO) to $gg$-fusion 
were computed and found to be large \cite{Dawson:1991zj, Djouadi:1991tk,
Spira:1995rr}. 
Since approximate evaluations \cite{Kramer:1998iq}
of higher-order terms suggest that their effect can still be sizeable,
the evaluation of the next-to-next-to-leading order (NNLO) corrections is 
highly desirable.

In this paper, we perform a first step towards the complete NNLO calculation.
We use the recently evaluated \cite{Harlander:2000mg}
two-loop amplitude for the process $gg \to H$ and the soft-gluon factorization
formulae \cite{Bern:1998sc, Catani:2000pi, Campbell:1998hg, Catani:2000ss}
for the bremsstrahlung subprocesses $gg \to Hg$ and $gg \to Hgg, Hq\bar{q}$, and we
compute the soft and virtual contributions to the NNLO partonic cross section.
We also discuss all-order resummation of soft-gluon contributions to
next-to-next-to-leading logarithmic (NNLL) accuracy.

We use the approximation $M_t \gg M_H$, where $M_t$ is the mass of the top
quark. The results of the NLO calculation in Ref.~\cite{Spira:1995rr} show that
this is a good numerical approximation \cite{Kramer:1998iq}
of the full NLO correction, provided 
the exact dependence on $M_H/M_t$ is included in the 
leading-order (LO) term. We can thus assume that the limit $M_t \gg M_H$
continues to be a good numerical approximation at NNLO.

The hadronic cross section for Higgs boson production is obtained by 
convoluting the perturbative partonic cross sections with the parton
distributions of the colliding hadrons. Besides the partonic cross sections, 
the other key ingredients of the NNLO calculation are
the NNLO parton distributions.
Even though their NNLO evolution kernels are
not fully available, some of their Mellin moments have been computed
\cite{vermaseren} and,
from these, approximated kernels have been constructed \cite{vnvogt}.
Recently, the new MRST \cite{mrst2000} sets of distributions became
available\footnote{We thank J. Stirling for providing us with the new set of
distributions.}, including the (approximated) NNLO 
densities, which allows an evaluation of the hadronic
cross section to (almost full) NNLO accuracy.

We use our NNLO result for the partonic cross sections and the 
MRST parton distributions at NNLO to compute the Higgs boson production
cross section at the LHC. In this paper, we do not present numerical results
for Run II at the Tevatron. Inclusive production of Higgs boson through
gluon--gluon fusion is phenomenologically less relevant at the Tevatron:
it is not regarded as a main discovery channel, because of the large QCD
background \cite{Carena:2000yx}.

The paper is organized as follows. In Sect.~\ref{sec:theo} we  define the
soft-virtual approximation for the cross section and present our result for
the corresponding NNLO coefficient. In  Sect.~\ref{sec:resummation} we discuss 
soft-gluon resummation for Higgs production at NNLL accuracy, and we also
consider the dominant contributions of collinear origin.
In Sect.~\ref{pheno} we present the quantitative effect of  
the computed NNLO corrections for SM Higgs boson production at the LHC.
Finally, in Sect.~\ref{sec:conc} we present our conclusions and
we comment on Higgs boson production beyond the SM.

\section{QCD cross section at NNLO}
\label{sec:theo}

We consider the collision of two hadrons $h_1$ and $h_2$ with 
centre-of-mass energy ${\sqrt s}$. The inclusive cross section for the 
production of the SM Higgs boson  can be written 
as
\begin{align}
\label{had}
\sigma(s,M_H^2) =& 
\sum_{a,b} \int_0^1 dx_1 \;dx_1 \; f_{a/h_1}(x_1,\mu_F^2) 
\;f_{b/h_2}(x_2,\mu_F^2) \int_0^1 dz \;\delta\!\left(z -
\frac{\tau}{x_1x_2}\right) \nn \\
& \cdot \sigma_0\,z\;G_{ab}(z;\as(\mu_R^2), M_H^2/\mu_R^2;M_H^2/\mu_F^2) \;,
\end{align}
where $\tau=M_H^2/s$, and $\mu_F$ and $\mu_R$ are factorization and 
renormalization scales, respectively. 
The parton densities of the colliding hadrons are denoted by 
$f_{a/h}(x,\mu_F^2)$ and the subscript $a$ labels the type
of massless partons ($a=g,q_f,{\bar q}_f$,
with $N_f$ different flavours of light quarks). 
We use parton densities as defined in the \ms\ factorization scheme.

From Eq.~(\ref{had}) the cross section ${\hat \sigma}_{ab}$ for the partonic 
subprocess $ab \to H + X$ at the centre-of-mass energy 
${\hat s}=x_1x_2s=M_H^2/z$ is
\begin{equation}
{\hat \sigma}_{ab}({\hat s},M_H^2) = \frac{1}{\hat s} 
\;\sigma_0 M_H^2 \;G_{ab}(z) = \sigma_0 \;z \;\;G_{ab}(z) \;,
\end{equation}
where the term $1/{\hat s}$ corresponds to the flux factor and leads to 
an overall $z$ factor. The  Born-level cross section $\sigma_0$ and 
the hard coefficient function $G_{ab}$ arise 
from the phase-space integral of the matrix elements squared. 

The incoming partons $a,b$ couple to the Higgs boson
through heavy-quark loops and, therefore,  $\sigma_0$ and $G_{ab}$ also depend on the 
masses $M_Q$ of the heavy quarks. The Born-level contribution $\sigma_0$
is~\cite{Georgi:1978gs}
\begin{equation}
\label{borncs}
\sigma_0=\f{G_F}{288\pi\sqrt{2}} 
\;| \sum_Q A_Q\!\left(\frac{4M_Q^2}{M_H^2}\right) |^2 \;\;,
\end{equation}
where $G_F=1.16639 \times 10^{-5}$ GeV$^{-2}$ is the Fermi constant,
and the amplitude $A_Q$ is given by
\begin{align}
A_Q(x)  &= \f{3}{2} x \Big[ 1+(1-x) f(x) 
\Big] \;,\nonumber \\
f(x) & = \left\{ \begin{array}{ll}
\displaystyle \arcsin^2 \frac{1}{\sqrt{x}} \;, & x \ge 1 \\
\displaystyle - \frac{1}{4} \left[ \ln \frac{1+\sqrt{1-x}}
{1-\sqrt{1-x}} - i\pi \right]^2 \;, & x < 1
\end{array} \right. \;.
\end{align}
In the following we limit ourselves to considering
the case of a single heavy quark, the top quark, and $N_f=5$ light-quark
flavours. We always use $M_t$ ($M_t=176$ GeV) 
 to denote the on-shell pole mass of the top quark.

The coefficient function $G_{ab}$ in Eq.~(\ref{had}) is computable in QCD
perturbation theory according to the expansion
\begin{align}
\label{expansion}
G_{ab}(z;\as(\mu_R^2), M_H^2/\mu_R^2;M_H^2/\mu_F^2) &=
\as^2(\mu_R^2) \sum_{n=0}^{+\infty} \left(\f{\as(\mu_R^2)}{\pi}\right)^n
\;G_{ab}^{(n)}(z;M_H^2/\mu_R^2;M_H^2/\mu_F^2) \nn \\
&= \as^2(\mu_R^2)\,
G_{ab}^{(0)}(z) + \f{\as^3(\mu_R^2)}{\pi} \;
G_{ab}^{(1)}\left(z;\frac{M_H^2}{\mu_R^2};\frac{M_H^2}{\mu_F^2}\right)\nn\\ 
&+ \f{\as^4(\mu_R^2)}{\pi^2} \;
G_{ab}^{(2)}\left(z;\frac{M_H^2}{\mu_R^2};\frac{M_H^2}{\mu_F^2}\right) 
+ {\cal O}(\as^5) \;,
\end{align}
where the (scale-independent) LO contribution is
\begin{equation}
G_{ab}^{(0)}(z) = \delta_{ag} \; \delta_{bg} \;\delta(1-z) \;.
\end{equation}

The NLO coefficients $G_{ab}^{(1)}$ are known. Their calculation with 
the exact dependence on $M_t$ was performed in Ref.~\cite{Spira:1995rr}.
In the large-$M_t$ limit (i.e. neglecting corrections that vanish
when  $M_H/M_t \to 0$) the result is \cite{Dawson:1991zj, Djouadi:1991tk}
\begin{align}
\label{gg1}
G_{gg}^{(1)}(z;M_H^2/\mu_R^2;&M_H^2/\mu_F^2)=
\delta(1-z) \left( \f{11}{2} + 6 \zeta(2) + \f{33-2N_f}{6}\ln\f{\mu_R^2}{\mu_F^2}\right)
+12\,\dl\nn\\
&+ 6\,\dO \,\ln\f{M_H^2}{\mu_F^2}\, 
+P_{gg}^{{\rm reg}}(z)\,\ln\f{(1-z)^2M_H^2}{z\mu_F^2}-6\f{\ln z}{1-z}
-\f{11}{2}\f{(1-z)^3}{z} \;,
\end{align}
\begin{equation}
\label{gq1}
G_{gq}^{(1)}(z;M_H^2/\mu_R^2;M_H^2/\mu^2_F)=\f{1}{2}P_{gq}(z)
\ln \f{(1-z)^2 M_H^2}{z\mu^2_F}+\f{2}{3}\, z-\f{(1-z)^2}{z} \;,
\end{equation}
\begin{equation}
\label{qq}
G_{q{\bar q}}^{(1)}(z;M_H^2/\mu_R^2;M_H^2/\mu_F^2)=\f{32}{27}\,\f{(1-z)^3}{z}
\;, \quad G_{qq}^{(1)}(z;M_H^2/\mu_R^2;M_H^2/\mu_F^2) = 0 \;,
\end{equation}
where $\zeta(n)$ is the Riemann zeta-function ($\zeta(2)=\pi^2/6=1.645\dots$,
$\zeta(3)=1.202\dots$), and we have defined
\begin{equation}
{\cal D}_i(z) \equiv \left[ \f{\ln^i(1-z)}{1-z}\right]_+\,\, .
\end{equation}
The kernels $P_{ab}(z)$ are the LO Altarelli--Parisi splitting functions
for real emission,
\begin{equation}
\label{apreg}
P_{gg}^{{\rm reg}}(z)=6\left[\f{1}{z}-2+z(1-z)\right] \;,~~~~~
P_{gq}(z)=\f{4}{3}\,\f{1+(1-z)^2}{z} \;,
\end{equation}
and, more precisely, $P_{gg}^{{\rm reg}}(z)$ is the regular part
(i.e. after subtracting the $1/(1-z)$ soft singularity) of $P_{gg}(z)$.
 
In Eqs.~(\ref{gg1})--(\ref{qq}) we can identify three kinds of contributions:
\begin{itemize}
\item Soft and virtual corrections, which involve only the $gg$ channel and
 give rise to the $\delta(1-z)$ and ${\cal D}_i$ terms in Eq.~(\ref{gg1}).
 These are the most singular terms when $z\to 1$.

\item Purely-collinear logarithmic contributions, which are controlled by 
the regular part of the Altarelli--Parisi splitting kernels 
(see Eqs.~(\ref{gg1}), (\ref{gq1})).
The argument of the collinear logarithm corresponds to the maximum value
$( q^2_{T \, {\rm max}} \sim (1-z)^2M_H^2/z)$ of the
transverse momentum $q_T$ of the Higgs boson. These contributions give the
next-to-dominant singular terms when $z\to 1$.

\item Hard contributions, which are present in all partonic channels and
lead to finite corrections in the limit $z\to 1$ .
\end{itemize}

The terms proportional to the distributions ${\cal D}_i(z)$  and 
$\delta(1-z)$ can be used to define what we call the
{\em soft-virtual} (SV) approximation. In this approximation only the 
$gg$ channel contributes and we have 
\begin{align}
\label{gg1SV}
G_{ab}^{(1) {\rm SV}}(z;M_H^2/\mu_R^2;M_H^2/\mu_F^2)&= \delta_{ag} \delta_{bg}
\left[ \delta(1-z) \left( \f{11}{2} + 6 \zeta(2) + 
\f{33-2N_f}{6}\ln\f{\mu_R^2}{\mu_F^2}\right) \right.\nn\\
&\left. +\,6\,\dO\, \ln\f{M_H^2}{\mu_F^2} +12\,\dl \right] \;.
\end{align}

The SV terms are certainly the dominant contributions to the cross section
in the kinematic region near threshold $(\tau = M_H^2/s \sim 1)$. At fixed
$s$, this means that the SV terms certainly dominate in the case of heavy Higgs
bosons. However, these terms 
can give the dominant effect even long before the
threshold region in the hadronic cross section is actually approached. This is
a consequence of the fact that the partonic cross section $\hat \sigma({\hat
s},M_H^2)$ has to be convoluted with the parton densities, and the QCD 
evolution of the latter sizeably reduces the energy that is available in the
partonic hard-scattering subprocess. Thus, the partonic cross section
$\hat \sigma({\hat s},M_H^2)$ (or the coefficient function 
$G(z)$) in the factorization formula (\ref{had}) is typically evaluated
much closer to threshold than the hadronic cross section. In other words,
the parton densities are large at small $x$ and are strongly suppressed
at large $x$ (typically, when $x \to 1$, $f(x, \mu^2) \sim (1-x)^\eta$ with
$\eta \gtap 3$ and $\eta \gtap 6$ for valence quarks and sea-quarks or gluons,
respectively); after integration over them, the dominant value of the square
of the partonic centre-of-mass energy 
$\langle {\hat s} \rangle = \langle x_1 x_2  \rangle s$ is therefore
substantially smaller than the corresponding hadronic value $s$. Note, also,
that this effect is enhanced, in gluon-dominated processes, by the
stronger suppression of the gluon density at large $x$.
In the case of Higgs boson production at the LHC, these features 
were emphasized in Ref.~\cite{Kramer:1998iq}, where the authors pointed out
that the SV approximation gives a good numerical approximation (see also
Sect.~\ref{pheno}) of the complete NLO corrections down to low values
($M_H \sim 100$~GeV) of the Higgs boson mass.

The NNLO coefficients $G_{ab}^{(2)}$ are not yet known. Their computation,
including their exact dependence on $M_t$, is certainly very difficult, since
it requires the evaluation of three-loop Feynman diagrams.

The computation is certainly more feasible in the large-$M_t$ limit, where
one can exploit the effective-lagrangian approach introduced in Ref.~\cite{efflag}
and developed up to ${\cal O}(\as^4)$ in Refs.~\cite{Chetyrkin:1997iv,
Kramer:1998iq}.
Using this approach, the contribution of the heavy-quark loop is
embodied by an effective vertex, thus reducing by one the number of loop
integrals to be explicitly carried out.

Within the effective-lagrangian formalism, an important step has recently
been performed by Harlander \cite{Harlander:2000mg}, who has evaluated the
two-loop amplitude for the process $gg\to H$ by using dimensional 
regularization in $d=4-2\epsilon$ space-time dimensions. 
The two-loop amplitude 
has poles of the type $1/\epsilon^n$ with $n=4,3,2,1$. The coefficients of the
poles of order $n=4,3,2$ had been predicted in Ref.~\cite{Catani:1998bh}. 
The agreement \cite{Harlander:2000ne} with this prediction is a non-trivial 
check of Harlander's result.

To compute the NNLO cross section, the {\em two-loop}
amplitude for the process $gg\to H$ has to be combined with the phase-space
integrals of the squares of the {\em one-loop} matrix element 
for the process $gg\to Hg$ and of the {\em tree-level} matrix elements
for the processes $gg\to Hgg$ and $gg\to Hq{\bar q}$. We have computed these
matrix elements in the limit where the final-state partons are soft, by using
the one-loop and tree-level factorization formulae derived in 
Refs.~\cite{Bern:1998sc,Catani:2000pi}
and Refs.~\cite{Campbell:1998hg,Catani:2000ss}, respectively.
Then, we have carried out the phase-space integrals by using the technique
of Ref.~\cite{Matsuura:1989sm}. The result contains $\epsilon$-poles and finite
terms. The $\epsilon$-poles (including the single pole $1/\epsilon$)
exactly cancel those in the two-loop amplitude \cite{Harlander:2000mg}, thus
providing a non-trivial cross-check of our and Harlander's results. 
The remaining
finite terms give the complete soft and virtual contributions to the NNLO
cross section.

Details of our calculation will be presented elsewhere \cite{inprep}.
In this paper we limit ourselves to presenting the final result. We obtain the
following soft and virtual contributions to the NNLO coefficient function
$G_{gg}^{(2)}$:
\begin{align}
\label{gg2SV}
G_{gg}&^{(2){\rm SV}}(z;M_H^2/\mu_R^2,M_H^2/\mu_F^2)= \delta(1-z)
\Bigg[ \f{11399}{144} +\f{133}{2}\zeta(2) -\f{9}{20}\zeta(2)^2
  -\f{165}{4}\zeta(3)\nn\\ 
&\hspace{4.5cm}
+  \left(\f{19}{8}+\f{2}{3} N_f\right)
 \ln\f{M_H^2}{M_t^2}
  +N_f \left( -\f{1189}{144} -\f{5}{3} \zeta(2) +\f{5}{6}\zeta(3)\right)
 \nn \\
&\hspace{4.5cm}
+\f{\left(33-2N_f\right)^2}{48}\ln^2\f{\mu_F^2}{\mu_R^2}- 18\,\zeta(2)
\ln^2\f{M_H^2}{\mu_F^2}\nn \\
&\hspace{4.5cm}
+ \left( \f{169}{4}+\f{171}{2}\zeta(3)- \f{19}{6}N_f
 +\left(33-2N_f\right)\zeta(2)\right)
\ln\f{M_H^2}{\mu_F^2}\nn \\
&\hspace{4.5cm}
+\left(-\f{465}{8}+\f{13}{3}N_f-\f{3}{2}\left(33-2N_f\right)\zeta(2)\right)
\ln\f{M_H^2}{\mu_R^2}
\Bigg] \nn \\
&
+\dO \Bigg[ -\f{101}{3} + 33 \zeta(2) + \f{351}{2} \zeta(3)+N_f\left(\f{14}{9}-2\zeta(2)\right)+
\left( \f{165}{4}-\f{5}{2}N_f \right)\ln^2\f{M_H^2}{\mu_F^2}\nn\\
&\hspace{1.2cm}
-\f{3}{2}\left(33-2N_f\right)\ln\f{M_H^2}{\mu_F^2}\,\ln\f{M_H^2}{\mu_R^2}  
+ \left( \f{133}{2} - 45 \zeta(2) - \f{5}{3} N_f \right)\ln\f{M_H^2}{\mu_F^2}
\Bigg] \nn \\
&
+\dl \Bigg[ 133 -90 \zeta(2) - \f{10}{3} N_f 
+ 36 \ln^2\f{M_H^2}{\mu_F^2}
+\left(33 -2 N_f \right)\left(2\ln\f{M_H^2}{\mu_F^2}-3\ln\f{M_H^2}{\mu_R^2}\right)
\Bigg]\nn \\
&+\dll \left[ -33+2 N_f +108 \ln\f{M_H^2}{\mu_F^2}\right]\nn\\
&+ 72\, \dlll \;\;.
\end{align}

Note that our result in Eq.~(\ref{gg2SV}) gives the complete soft contributions
(all the
terms proportional to the distributions ${\cal D}_i(z)$)
to the NNLO coefficient functions $G_{ab}^{(2)}(z)$.
It also gives the complete virtual contribution
(the term proportional to $\delta(1-z)$)
to the $gg$ channel. The expression in 
Eq.~(\ref{gg2SV}) is an approximation of the exact NNLO calculation in the 
sense that it differs from $G_{ab}^{(2)}(z)$ by terms that are less singular
when $z \to 1$. More precisely, in the large-$z$ limit
we have (see Sect.~(\ref{sec:resummation}))
\begin{align}
&G_{gg}^{(2)}(z) - G_{gg}^{(2){\rm SV}}(z) = {\cal O}( \ln^3(1-z)) \;\;, \\
\label{gq2}
&G_{gq}^{(2)}(z) = {\cal O}( \ln^3(1-z)) \;\;,\\
&G_{q{\bar q}}^{(2)}(z) \propto  \delta(1-z) + {\cal O}((1-z) \ln^2(1-z)) \;\;,
\quad G_{qq}^{(2)}(z) = {\cal O}((1-z) \ln^2(1-z))
\;\;.
\end{align}

Note also that, unlike the NLO term $G_{ab}^{(1)}(z)$,
the NNLO coefficient function $G_{ab}^{(2)}(z)$ is not independent of $M_t$ in 
the large-$M_t$ limit. The virtual contribution in Eq.~(\ref{gg2SV}) contains a term, 
proportional to $\ln M_H^2/M_t^2$, that derives from the integration of the
heavy-quark degrees of freedom in the effective lagrangian 
\cite{Chetyrkin:1997iv, Kramer:1998iq}.

Our result in Eq.~(\ref{gg2SV}) can be useful as a non-trivial check
of a future complete calculation at NNLO. It can also be used to extend the
accuracy of the soft-gluon resummation formalism to NNLL order (see 
Sect.~\ref{sec:resummation}).

As previously discussed, the SV approximation turns out to be a good numerical
approximation of the full NLO correction for Higgs boson production at the LHC.
Thus, the NNLO-SV result in Eq.~(\ref{gg2SV}) can also be exploited to
obtain an approximate numerical estimate of the complete NNLO correction
(see Sect.~\ref{pheno}).
 
\section{Soft-gluon
resummation at NNLL accuracy}
\label{sec:resummation}

The soft (and virtual) contributions $\as^2 \as^n {\cal D}_m(z)$ (with $m \leq
2n-1$) to the coefficient function $G_{gg}(z)$ can be summed to all orders in
QCD perturbation theory. Using the soft-gluon resummation formulae that
are known at present,
we can check the coefficients of some of the soft contributions presented in 
Eq.~(\ref{gg2SV}). The remaining coefficients can then be used to extend
the accuracy of the resummation formulae to NNLL order. Both points are
discussed in this section.

The formalism to systematically perform soft-gluon resummation for processes
initiated by $q{\bar q}$ annihilation and $gg$ fusion was set up in 
Refs.~\cite{Sterman, CT, Catani:1988vd, Catani:1991rr}. Soft-gluon resummation
has to be carried out in the Mellin (or $N$-moment) space. The $N$-moments
$G_N$ of the coefficient function $G(z)$ are defined by
\begin{equation}
G_N \equiv \int_0^1 dz \;z^{N-1} \;G(z) \;\;.
\end{equation}
In $N$-moment space the soft (or threshold) region $z \to 1$ 
corresponds to the limit $N\to \infty$, and the distributions ${\cal D}_m(z)$
lead to logarithmic contributions, ${\cal D}_m(z) \to \ln^{m+1}N$. The singular
contributions in the large-$N$ limit can be organized in the following
{\em all-order} resummation formula:
\begin{equation}
\label{resformulaG}
G_{gg, \,N} = {\overline C}_{gg}(\as(\mu^2_R),M_H^2/\mu^2_R;M_H^2/\mu_F^2) 
\; \Delta_{N}^{H}(\as(\mu^2_R),M_H^2/\mu^2_R;M_H^2/\mu_F^2)\; + {\cal O}(1/N)
\;\;. 
\end{equation}
The radiative factor $\Delta_N^{H}$ embodies all the large contributions $\ln N$
due to soft radiation.
The function ${\overline C}_{gg}(\as)$ contains all the terms that are
constant in the large-$N$ limit and has a perturbative expansion analogous 
to Eq.~(\ref{expansion}):
\begin{align}
\label{cbarexp}
{\overline C}_{gg}(\as(\mu^2_R),&M_H^2/\mu^2_R;M_H^2/\mu_F^2) =\nn\\
&=\as^2(\mu^2_R)
\left[ 1 + \sum_{n=1}^{+\infty} \; 
\left( \frac{\as(\mu^2_R)}{\pi} \right)^n \;
{\overline C}_{gg}^{(n)}(M_H^2/\mu^2_R;M_H^2/\mu_F^2) \right] \;.
\end{align}
These constant terms are due to virtual contributions, and 
the perturbative coefficients ${\overline C}_{gg}^{(n)}$ are thus directly 
related to the coefficients of the contribution proportional to
$\delta(1-z)$ in $G_{gg}^{(n)}(z)$. The term ${\cal O}(1/N)$ on the right-hand
side of Eq.~(\ref{resformulaG}) denotes all the contributions that are
suppressed by some power of $1/N$ (modulo $\ln N$ enhancement) when 
$N \to \infty$.

The radiative factor $\Delta_N^{H}$ for Higgs boson production
has the following general expression
\cite{Sterman, CT, Catani:1998tm}:
\begin{align}
\label{deltan}
\Delta_N^{H}(\as(\mu^2_R),M_H^2/\mu^2_R;M_H^2/\mu_F^2) &=
\left[ \Delta_N^g(\as(\mu^2_R),M_H^2/\mu^2_R;M_H^2/\mu_F^2) \right]^2\nn\\
&\cdot \Delta_N^{({\rm int})  H}(\as(\mu^2_R),M_H^2/\mu^2_R) \;\;.
\end{align}
Each term $\Delta_N^g$ embodies
the effect of soft-gluon radiation emitted collinearly to the initial-state 
partons and depends on both the factorization 
scheme and the factorization scale $\mu_F$. In the $\msbar$ factorization
scheme we have the {\em exponentiated} result
\begin{equation}
\label{deltams}
\Delta_N^a(\as(\mu^2_R),M_H^2/\mu^2_R;M_H^2/\mu_F^2) = \exp \Big\{ 
\int_0^1 dz \;\frac{z^{N-1} -1}{1-z} \;
\int_{\mu_F^2}^{(1-z)^2M_H^2} 
\frac{dq^2}{q^2} \;A_a(\as(q^2))  \Big\} \;\;,
\end{equation}
where $A_a(\as)$ is a perturbative function
\beeq
\label{Afun}
A_a(\as)={\asp} A_a^{(1)}+\left(\asp \right)^2 A_a^{(2)} 
+\left(\asp \right)^3 A_a^{(3)}+ {\cal O}(\as^4)
\;\;.
\eeeq 
The factor $\Delta_N^{({\rm int})}$ is independent of
the factorization scale and scheme and contains the 
contribution of soft-gluon emission at large angles with respect to the
direction of the colliding gluons. It can also be written in exponentiated form
as
\begin{equation}
\label{deltaint}
\Delta_N^{({\rm int})  H}(\as(\mu^2_R),M_H^2/\mu^2_R) = \exp \Big\{
\int_0^1 dz  \;\frac{z^{N-1} -1}{1-z} \; 
\;D_{H}(\as((1-z)^2M_H^2)) \Big\}\;,
\end{equation}
where the function $D_{H}(\as)$ for Higgs production
has the following perturbative expansion: 
\beeq
\label{Dfun}
D_{H}(\as) = \left( \f{\as}{\pi} \right)^2 D_{H}^{(2)}
 + {\cal O}(\as^3)\;.
\eeeq

The coefficients $A^{(1)}$ and $A^{(2)}$ fully control soft-gluon resummation
up to next-to-leading logarithmic (NLL) accuracy 
\cite{Sterman, CT, Catani:1998tm}. In the case of a generic incoming parton 
$a$, they are given by
\beeq
\label{A12coef}
A_a^{(1)}= C_a\;,\;\;\;\; A_a^{(2)}=\frac{1}{2} \; C_a K \;,
\eeeq
where $C_a = C_F =4/3$ if $a=q,{\bar q}$ and $C_a = C_A=3$ if $a=g$, while
the coefficient $K$ is the same both for quarks and for gluons 
\cite{KT,Catani:1988vd,deFlorian:2000pr} and it is given by
\beeq
\label{kcoef}
K = C_A \left( \frac{67}{18} - \frac{\pi^2}{6} \right) 
- \frac{5}{9} N_f \;.
\eeeq
Expanding the resummation formula (\ref{resformulaG}) up to 
${\cal O}(\as^3)$ and transforming the result back to $z$-space, it is
straightforward to check that we correctly obtain the soft contributions,
${\cal D}_0(z)$ and ${\cal D}_1(z)$, 
to $G_{ab}^{(1){\rm SV}}(z)$ in Eq.~(\ref{gg1SV}). By comparison with
the virtual term in Eq.~(\ref{gg1SV}), we can also extract the coefficient 
${\overline C}_{gg}^{(1)}$ in Eq.~(\ref{cbarexp}):
\begin{align}
\label{cbar1}
{\overline C}_{gg}^{(1)}(M_H^2/\mu^2_R;M_H^2/\mu_F^2)= 
\f{11}{2} + 6 \zeta(2) + \f{33-2N_f}{6}\ln\f{\mu_R^2}{\mu_F^2} \;\;.
\end{align}
Then, we can expand the resummation formula (\ref{resformulaG}) up to 
${\cal O}(\as^4)$, and we can compare the result with our NNLO soft-virtual
calculation in Eq.~(\ref{gg2SV}). It is straightforward to check that
the knowledge of $A_g^{(1)}, A_g^{(2)}$ and ${\overline C}_{gg}^{(1)}$
predicts the coefficients of ${\cal D}_3(z), {\cal D}_2(z)$ and 
${\cal D}_1(z)$ in $G_{gg}^{(2){\rm SV}}(z)$, and that the prediction fully
agrees with our result in Eq.~(\ref{gg2SV}).

The comparison\footnote{We can also extract the virtual coefficient
${\overline C}_{gg}^{(2)}$ in Eq.~(\ref{cbarexp}).}
at ${\cal O}(\as^4)$ and our calculation
of the ${\cal D}_0\,$-term in Eq.~(\ref{gg2SV})
also allows us to extract the 
(so far unknown) coefficient $D_H^{(2)}$ that controls soft-gluon resummation
at NNLL order. We obtain
\begin{align}
D_H^{(2)}= C_A^2 
\left( -\f{101}{27} + \f{11}{3} \zeta(2)+ \f{7}{2} \zeta(3) \right)+ 
C_A N_f \left( \f{14}{27} -\f{2}{3}\zeta(2)\right) \;\;.
\end{align}
Note that the corresponding NNLL coefficient
for the Drell--Yan process \cite{vogtresum} differs from $D_H^{(2)}$
by the simple replacement of colour factors $C_F \to C_A$.
This could have straightforwardly been predicted from the general structure
of the soft-factorization formulae at ${\cal O}({\as^2})$ (see Sect.~5 of 
Ref.~\cite{Catani:2000pi} and the Appendix of Ref.~\cite{Catani:2000ss}).
The exact expression of the remaining NNLL coefficient $A_g^{(3)}$ is still
unknown, but an approximate numerical estimate can be found in 
Ref.~\cite{vogtresum}.

The integrals over $z$ and $q^2$ in Eqs.~(\ref{deltams}) and (\ref{deltaint})
can be carried out to any required logarithmic accuracy
(see Refs.~\cite{Catani:1998tm, vogtresum}) and used for phenomenological
analyses.  Quantitative studies of soft-gluon resummation effects for Higgs
boson production are left to future investigations.

\subsection{Collinear-improved resummation}
\label{sec:svc}

In Ref.~\cite{Kramer:1998iq} Kr\"amer, Laenen and Spira (KLS) exploited 
the resummation formalism to obtain approximate expressions for the NNLO
corrections to Higgs boson production. Their resummation formula is a 
simplified version of Eq.~(\ref{resformulaG}) that includes only the
first-order coefficients (the coefficients $A^{(1)}, {\overline C}^{(1)}$ and 
the first-order coefficient $\beta_0$ in the expression of the running coupling
$\as(q^2)$). Therefore, the NNLO expressions obtained in 
Ref.~\cite{Kramer:1998iq} correctly predict only the coefficients of the
contributions ${\cal D}_3$ and ${\cal D}_2$ to the soft and virtual coefficient
function $G_{gg}^{(2){\rm SV}}$ in Eq.~(\ref{gg2SV}). 

KLS also pointed out \cite{Kramer:1998iq} that the resummation formalism can 
be extended to include subdominant contributions in the large-$z$ 
limit. These contributions are the terms proportional to powers of 
$\ln (1-z)$ that appear in $G_{gg}(z)$ (see, e.g., Eq.~(\ref{gg1})). In
$N$-moment space, they lead to contributions of the type $\frac{1}{N} \ln^kN$,
which are usually (and consistently) neglected within the soft and virtual 
approximation (i.e. in the limit $N \to \infty$). 

We agree with KLS that
the {\em highest} power\footnote{As for lower powers, KLS acknowledge 
\cite{Kramer:1998iq} that their result is not complete.} 
of $\ln (1-z)$ at the $n$-th perturbative order,
namely, $\ln^{2n-1}(1-z)$ in $G_{gg}^{(n)}(z)$ (or, equivalently, the term 
$\frac{1}{N} \ln^{2n-1}N$ in $G_{gg, \,N}^{(n)}$), can correctly and 
consistently be implemented in the all-order resummation formula 
(\ref{resformulaG}). The key observation \cite{Kramer:1998iq} is that these
terms have a {\em collinear} origin. They arise from the transverse-momentum
evolution of initial-state collinear radiation up to the maximum value of
$q_T$ permitted by kinematics. In the large-$z$ limit, the maximum value
is $q^2_{T \, {\rm max}} \sim (1-z)^2M_H^2$, which is very different from
the typical hard scale $M_H^2$ of the process. The large transverse-momentum
region $(1-z)^2M_H^2 < q^2_T < M_H^2$ is thus responsible for the leading
$\ln(1-z)$-enhancement. The resummation formalism correctly
embodies the transverse-momentum evolution of soft radiation up to the
kinematical limit $(1-z)^2M_H^2$ (see Eq.~(\ref{deltams})). Therefore,
the leading collinear enhancement can be taken into account by
supplementing the integrand in Eq.~(\ref{deltams}) with the regular
(i.e. non-soft) part of the Altarelli--Parisi splitting function (see
Eq.~(\ref{apreg})). Both for the $q{\bar q}$ annihilation (Drell--Yan process)
and $gg$ fusion (Higgs production) channels,
we can simply perform the following replacement on the right-hand side
of Eq.~(\ref{deltams}):
\begin{align}
\label{replac}
\frac{z^{N-1} -1}{1-z} \;A_a^{(1)}  &\to\;
\frac{z^{N-1} -1}{1-z} \;A_a^{(1)} + z^{N-1} \,\frac{1}{2} P_{aa}^{{\rm reg}}(z)=\nn\\ 
&= \left[ \frac{z^{N-1} -1}{1-z} - z^{N-1} \right] A_a^{(1)} + {\cal O}(1/N^2)
\;\;.
\end{align}

Having performed the replacement of Eq.~(\ref{replac})
in $\Delta_N^a$, we can insert its ensuing
{\em collinear-improved} expression in Eq.~(\ref{resformulaG}). The resummed
expression for the $N$ moments of the coefficient function $G_{gg, \,N}$
can then be expanded in powers of $\as$ in the large-$N$ limit by consistently
computing and keeping all the terms of the type $\as^2 \as^n \frac{1}{N}
\ln^{2n-1}N$. Transforming the result back to $z$-space, this procedure gives
the soft and virtual contributions to $G_{gg}^{(n)}(z)$ plus its
{\em leading} subdominant correction (the contribution proportional to 
$\ln^{2n-1}(1-z)$) when $z \to 1$. 

We name {\em soft-virtual-collinear} (SVC) 
approximation this improved version of the SV expressions in Eqs.~(\ref{gg1SV})
and (\ref{gg2SV}). We find
\begin{equation}
\label{gg1SVC}
 G_{gg}^{(1){\rm SVC}}(z;M_H^2/\mu^2_R;M_H^2/\mu^2_F) =
G_{gg}^{(1){\rm SV}}(z;M_H^2/\mu^2_R;M_H^2/\mu^2_F) -12\, 
\ln(1-z) \;\;,
\end{equation}
\begin{equation}
\label{gg2SVC}
G_{gg}^{(2){\rm SVC}}(z;M_H^2/\mu^2_R;M_H^2/\mu^2_F) =
G_{gg}^{(2){\rm SV}}(z;M_H^2/\mu^2_R;M_H^2/\mu^2_F) - 72\, \ln^3(1-z) \;\;.
\end{equation}
The coefficient of $\ln(1-z)$ in Eq.~(\ref{gg1SVC}) correctly reproduces that 
obtained by the exact NLO expression in Eq.~(\ref{gg1}). The coefficient of 
$\ln^3(1-z)$ in Eq.~(\ref{gg2SVC}) agrees with that computed in 
Ref.~\cite{Kramer:1998iq}.

The numerical study of Ref.~\cite{Kramer:1998iq} shows that the effect of the
contribution $\ln(1-z)$ at NLO is not small (see also Sect.~\ref{pheno}), 
in particular at low values of the Higgs boson mass.
Therefore, in our estimate (Sect.~\ref{pheno})
of the NNLO corrections to Higgs boson production at the LHC,
we consider both the SV approximation in Eq.~(\ref{gg2SV}) and
the SVC approximation in Eq.~(\ref{gg2SVC}). In the $gg$ partonic channel
we thus neglect NNLO of the type
\begin{equation}
G_{gg}^{(2)}(z) - G_{gg}^{(2){\rm SVC}}(z) = {\cal O}( \ln^2(1-z)) \;\;,
\end{equation}

Note, however, that the coefficient function of the $gq$ channel still
contains contributions proportional to $\ln(1-z)$ at NLO (see Eq.~(\ref{gq1}))
and to $\ln^3(1-z)$ at NNLO (see Eq.~(\ref{gq2})). We do not consider the
latter. At low values of the Higgs boson mass their effect is small, because
the parton density luminosity of the $gq$ channel is smaller than that of
the $gg$ channel. The effect increases by increasing the Higgs boson mass. 

\section{Numerical results at the LHC}

\label{pheno}
In this section we
study
the phenomenological impact of the higher-order QCD
corrections on the production of the SM Higgs boson at the LHC,
 i.e. proton--proton collisions at
$\sqrt{s}=14$~TeV. 
We recall that we include the exact dependence on $M_t$ in the Born-level 
cross section $\sigma_0$ (see Eq.~(\ref{borncs})), while the coefficient 
function $G_{ab}(z)$ is evaluated in the large-$M_t$ approximation. At NLO 
\cite{Spira:1995rr, Kramer:1998iq}
this is a very good numerical approximation when $M_H \leq 2M_t$, and it is 
still accurate to better than $10\%$ when $M_H \ltap 1$~TeV.

 Unless otherwise stated, cross sections are computed
 using the new  MRST2000 \cite{mrst2000} sets of
 parton distributions, with densities and coupling constant evaluated at each
 corresponding order, i.e. using LO distributions 
and 1-loop $\as$ for the LO
 cross section, and so forth. The corresponding values of
 $\Lambda^{(5)}_{QCD}$ ($\as(M_Z)$)  are  $0.132$ (0.1253), $0.22$ (0.1175)
   and $0.187$ GeV (0.1161), at 1-loop, 2-loop and 3-loop order,
   respectively.
 In the  NNLO case  we  use the `central' set of
 MRST2000, obtained from a global fit of data (deep inelastic scattering,
 Drell--Yan production and jet $E_T$ distribution) by using the approximate
 NNLO evolution kernels presented in Ref.~\cite{vnvogt}. 
The result we refer to as NNLO-SV (SVC) corresponds to the sum of the
LO and exact NLO (including the $qg$ and $q\bar{q}$ channels)
contributions plus the SV (SVC) corrections at NNLO, given in Eq.~(\ref{gg2SV}) 
(Eq.~(\ref{gg2SVC})).
The LO and NLO results obtained by using the CTEQ5 
 distributions \cite{cteq5} are very similar to the ones computed with 
 the MRST2000 sets
(the differences are smaller than the uncertainties arising, for instance,
 from scale dependence). Therefore, we will not show those
 results \footnote{Larger deviations (for instance, the NLO cross section increases by $\sim 10\%$ for $M_H=100-200$ GeV) appear when comparing to
   the GRV98 distributions \cite{Gluck:1998xa}, where both the gluon
distribution and the value of $\as(M_Z)$ are different from those of MRST2000 and CTEQ5.}.

The comparison between different sets of parton distributions, however, cannot
be regarded as a way to quantitatively estimate the uncertainty on the 
parton distributions. The theoretical and experimental errors that affect
present determinations of the parton distributions are typically larger 
\cite{Catani:2000jh} than the differences between the parton distribution
sets provided by different groups 
\cite{mrst2000, cteq5, Gluck:1998xa}. In the case of Higgs boson 
production at the LHC, the study of the CTEQ Collaboration 
\cite{Huston:1998jj} recommends an
uncertainty of about $\pm 10\%$ on the corresponding gluon--gluon and
quark--gluon parton luminosities.

\begin{figure}[htb]
\begin{center}
\begin{tabular}{c}
\epsfxsize=15truecm
\epsffile{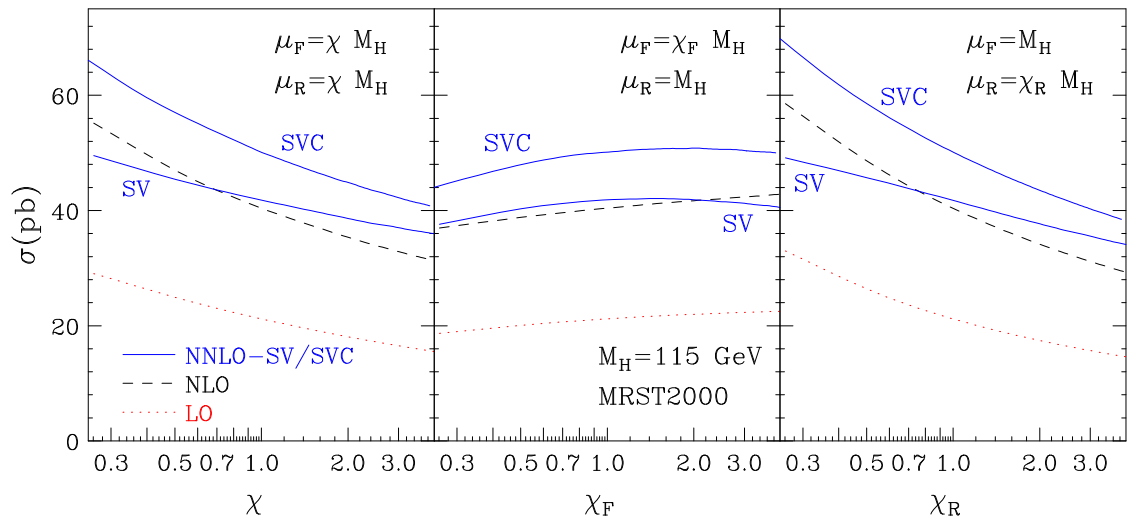}\\
\end{tabular}
\end{center}
\caption{\label{fig:scale}{\em Scale dependence of the Higgs production cross section for
    $M_H=115$ GeV at LO, NLO, NNLO-SV and NNLO-SVC. }}
\end{figure}

We begin the presentation of our results by showing in Fig.~1
the scale dependence of the cross section for the production of a Higgs boson
with $M_H=115$ GeV. The scale dependence is analysed by varying the factorization
 and renormalization scales by a factor of 4 up and down from the default value $M_H$. 
The plot on the left corresponds to the simultaneous variation of both scales,
 $\mu_F=\mu_R=\chi \, M_H$,
whereas the plots in the centre and on the right correspond, respectively, to 
the results of the independent variation of the factorization or 
renormalization scale, keeping the other scale fixed at the default value.

As expected from the QCD running of $\as$, the cross sections typically 
decrease when $\mu_R$ increases  around the characteristic hard scale $M_H$.
In the case of variations of $\mu_F$, we observe the opposite behaviour. In
fact, the cross sections are mainly sensitive to partons with momentum fraction
$x \sim 10^{-2}$, and in this $x$-range scaling violation of the parton
densities is (moderately) positive. As a result, the scale dependence is 
mostly driven by the renormalization scale, because the lowest-order 
contribution  to the process is proportional to $\as^2$, 
a (relatively) high power of $\as$.

Figure ~1 shows that
the scale dependence is reduced when higher-order corrections are included and, in the case of the 
factorization-scale dependence, a maximum appears at NNLO-SV and NNLO-SVC,
 showing the
improved stability of the result. Also note that
there is an increase in the scale dependence when going from NNLO-SV to NNLO-SVC. 
This is due to the fact that the dominant collinear terms included in the SVC
approximation give a sizeable contribution and are scale-independent 
(see Eqs.~(\ref{gg1SVC}) and (\ref{gg2SVC})), so their effect cannot be
compensated by scale variations.
Similar results are obtained for higher masses, with a  reduction in the scale
 dependence when approaching high mass values.

\begin{figure}[htb]
\begin{center}
\begin{tabular}{c}
\epsfxsize=12truecm
\epsffile{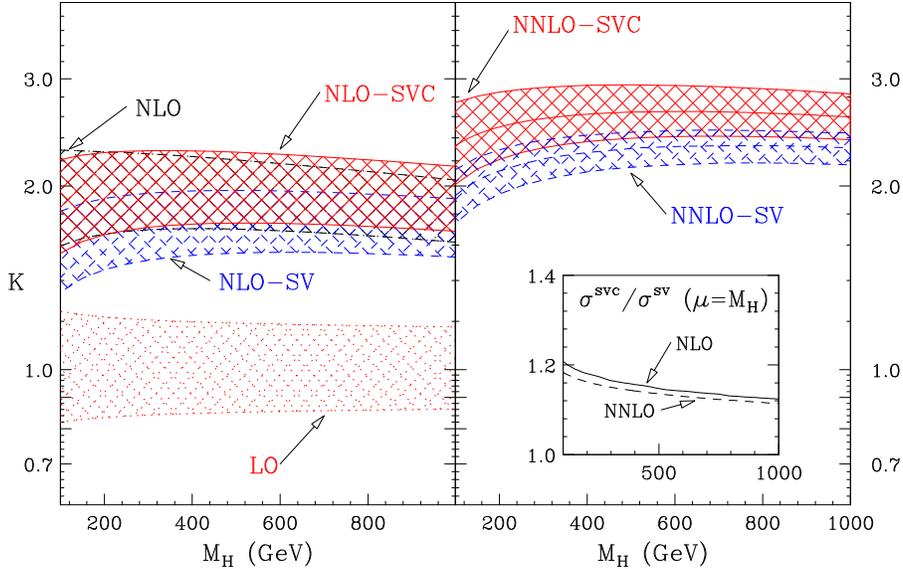}\\
\end{tabular}
\end{center}
\caption{\label{fig:k1}{\em  K-factors for Higgs production for the full NLO result and the NLO-SV, NLO-SVC, NNLO-SV and NNLO-SVC approximations.}}
\end{figure}

The impact of higher-order corrections is usually studied by computing
$K$-factors, defined as the ratio of the cross section evaluated at
each corresponding order over the LO result.
The $K$-factors are shown in Fig.~2 where the bands account for the 
`theoretical uncertainty' due to 
the scale dependence, quantified by using the  minimum and maximum values of 
the cross sections when the scales $\mu_R$ and $\mu_F$
are varied (simultaneously and independently, as in Fig.~\ref{fig:scale})
 in the range $0.5 \le \chi, \chi_R, \chi_F\le 2$. The LO result that 
 normalizes the $K$-factors is computed at the default scale $M_H$ in all cases.

The plot on the left-hand side of Fig.~\ref{fig:k1} shows
the uncertainty at LO and compares the exact NLO result with the NLO-SV and
 NLO-SVC approximations. In the case of light Higgs production, the NLO-SV 
 approximation  
 tends to underestimate the exact result by about $15$ to $20\%$, whereas the 
 NLO-SVC approximation
 only slightly overestimates it, showing the numerical importance of the term 
 $\ln(1-z)$ added in the SVC approximation. Nevertheless, all the 
results agree within the theoretical bands: 
this  
confirms the validity of the large-$z$ approximation to estimate higher-order 
corrections, and, in particular, allows us to assume that a similar situation
occurs at NNLO.
As expected, the agreement between the three results improves for larger 
masses. 

The right-hand side of Fig.~\ref{fig:k1} shows the SV and SVC results at NNLO. Again, 
the SVC band
sits higher
than the SV one, the ratio of the corresponding cross sections 
being almost the same as the one at NLO, as shown in the inset plot.
The contribution from non-leading terms $\ln^k(1-z)$, with $k<3$
(which are not under control within the SVC approximation), is not included,
but it is expected to be numerically less important\footnote{We have tried
to add a term $\ln^2(1-z)$ with a coefficient as large as that of the term
$\ln^3(1-z)$, finding  only a small (about $5\%$) modification.}.

As is well known, the customary procedure (that we also are using) of varying
the scales to estimate the theoretical uncertainty can only give a lower limit
on the `true' uncertainty. This is well demonstrated by Fig.~\ref{fig:k1}, which
shows no overlap between the LO and NLO bands. However, the NLO and NNLO bands
do overlap, thus suggesting that the perturbative expansion begins to converge
from NNLO. Note also that the size of the NNLO bands is smaller than that of 
the NLO bands: the scale dependence at NNLO is smaller than at NLO.

 Considering the results obtained at NLO, it is reasonable to expect the 
full NNLO $K$-factor
to lie inside the SV and SVC bands, and most probably, 
closer to the SVC one. 
In particular, for a light Higgs boson ($M_H \ltap 200$~GeV), this expectation would correspond to an increase of
 15 to 25$\%$ with respect to the full NLO result, i.e. a factor of about 2.2 to 2.4 with respect to 
the LO result.
Taking into account that the NLO result increases the
 LO cross section by about 90$\%$ our result anticipates a good convergence of the perturbative series.
\begin{figure}[htb]
\begin{center}
\begin{tabular}{c}
\epsfxsize=8truecm
\epsffile{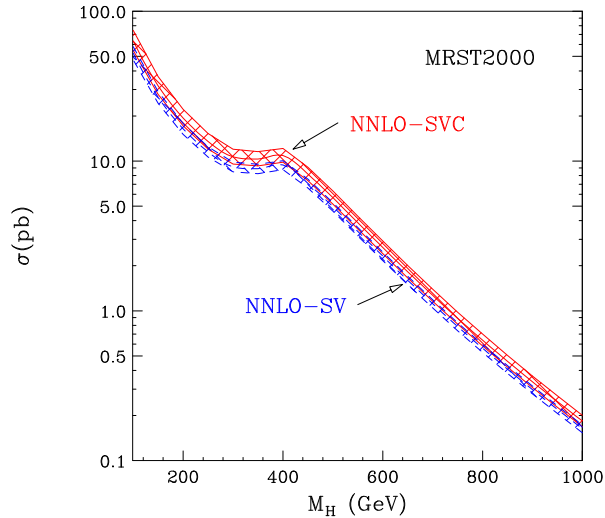}\\
\end{tabular}
\end{center}
\caption{\label{fig:xsec}{\em Cross section for Higgs boson production at the LHC in the NNLO-SV and NNLO-SVC approximations.}}
\end{figure}

In Fig.~3 we present the NNLO-SV and SVC cross sections as a function of the 
Higgs mass and including the corresponding uncertainty bands computed as 
defined above. To facilitate the comparison with other calculations 
and more refined predictions, we report the values of the cross sections
for the production of a Higgs boson with $M_H=115$ GeV. The NNLO-SVC band
corresponds to 
$\sigma= 43.51$-58.56~pb (50.13~pb at the default scales),
 the NNLO-SV to
$\sigma= 37.73$-45.69~pb (41.66~pb at the default scales), whereas for the full NLO it is
$\sigma= 34.14$-48.48~pb (40.37~pb at the default scales).

\begin{figure}
\begin{center}
\begin{tabular}{c}
\epsfxsize=11truecm
\epsffile{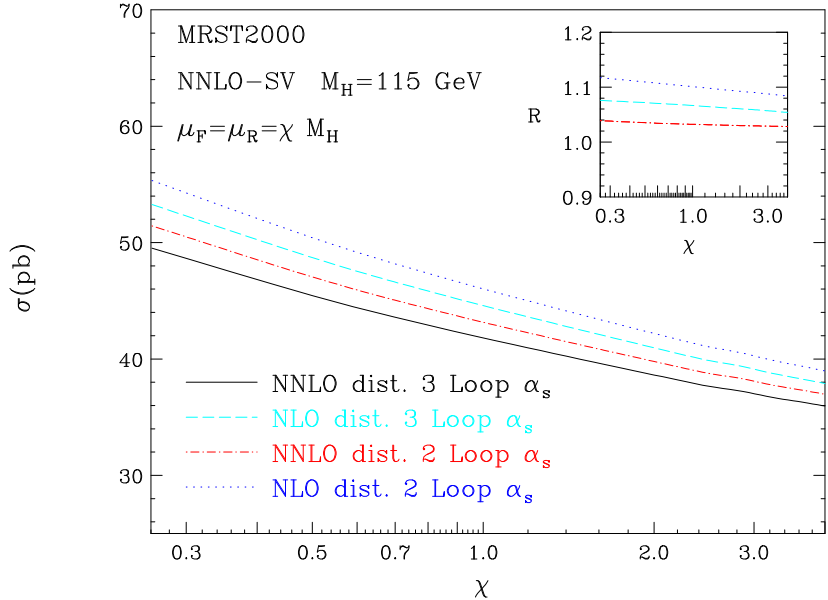}\\
\end{tabular}
\end{center}
\caption{\label{fig:dist}{\em Cross section for Higgs production with $M_H=115$ GeV computed using different NLO and NNLO parton distributions and coupling constant.}}
\end{figure}

Finally
we want to
quantify  the effect of
 the (approximated) NNLO parton distributions  in the gluonic channel.
In Fig.~4 we study this effect
for the NNLO-SV result at $M_H=115$ GeV, by plotting the cross section 
as a function of the scale. We use 
different combinations of NNLO and NLO parton distributions 
and coupling constant expressions. The inset plot shows the ratio $R$ of
 the different results with respect to the one obtained by using NNLO 
 distributions and 3-loop $\as$.
The use of 
NNLO distributions and 3-loop $\as$ 
reduces the NNLO cross section by $10 \%$
with respect to the result that would be obtained if NLO distributions and
 2 loop $\as$ were used. Since the values of $\as(M_Z)$ from MRST2000 are very
 similar at 2 and 3 loops and the typical scale of the process is not far from
 $M_Z$, the  effect of going from 2 to 3 loops $\as$ amounts to only $1/3$ of the
 $ 10 \%$ change. The biggest effect comes from the difference in the
 distributions, mostly due to the decrease of the NNLO gluon density at small
 $x$ \cite{mrst2000}. Similar results are obtained in the SVC approximation
 and for different  masses. 

\section{Conclusions}
\label{sec:conc}

In this paper we have studied the QCD corrections to Higgs boson production 
through gluon--gluon fusion in hadronic collisions, within the framework of 
the large-$M_t$ approximation. Using a recent result for the two-loop correction to the $gg\to H$ amplitude \cite{Harlander:2000mg} and the soft-factorization formulae for soft-gluon emission at ${\cal O}(\as^2)$
\cite{Bern:1998sc,Catani:2000pi,Campbell:1998hg,Catani:2000ss}
we have evaluated the soft and virtual QCD correction to this process 
at NNLO (SV approximation). 
We have also considered \cite{Kramer:1998iq} 
the leading $\ln^3(1-z)$ contribution from the collinear region (SVC approximation).
Our result for the coefficient $G_{gg}^{(2)SV}$ in Eq.~(\ref{gg2SV}) 
is consistent with the present knowledge of soft-gluon resummation at 
NLL accuracy; it also allows us to fix the NNLL coefficient $D^{(2)}_H$ in 
Eq.~(\ref{Dfun}).

We have then studied the phenomenological impact of our results at the LHC by using
the (approximate) NNLO set MRST2000 of parton distributions \cite{mrst2000}.
We have shown that the exact NLO result lies in between the NLO-SV 
approximation and the NLO-SVC approximation, the latter being a better numerical
approximation in the case of low values
of the Higgs boson mass.
Comparing the results in the SV and SVC approximations at NNLO for 
a light Higgs ($M_H \ltap 200$~GeV), we estimate that
the NNLO correction will increase the NLO result between $15$ and $25\%$.

The results presented here are a first consistent (though approximate)
estimate of QCD corrections to Higgs boson production through $gg$ fusion at NNLO and will eventually be a stringent check of a future full NNLO calculation.

In this paper we have only considered the production of the SM Higgs boson.
The Minimal Supersymmetric extension of the Standard Model (MSSM) leads to two
CP-even neutral Higgs bosons \cite{Gunion:1989we}. They are produced 
by $gg$ fusion through loops of heavy quarks (top, bottom) and squarks.
For small values $(\tan \beta \ltap 5)$ of the MSSM parameter $\tan \beta$,
the NLO QCD corrections to this production mechanism are comparable (to better
than $10\%$) \cite{Spira:1995rr, DDS, Spira:1998dg} to those for SM Higgs 
boson production. Therefore, the NNLO
$K$-factors computed in this paper could also be applicable to MSSM
Higgs boson production.

{\bf Note added:} The calculation of the soft and virtual NNLO corrections to Higgs boson production has independently been performed in Ref.\cite{HK}. The method used in Ref.\cite{HK} is different from ours. The analytical results fully agree.

\end{document}